\documentclass[twocolumn, a4paper, aps, pra, 10pt]{revtex4-1}

  \usepackage[utf8]{inputenc}                  % Eingabecodierung wird auf unicode utf8 gesetzt
  \usepackage[english]{babel}                  % deutsche Sprache mit neuer Rechtschreibung
  \usepackage[T1]{fontenc}                     % Codierung der Schriftart
  \usepackage{lmodern}                         % Setze latin modern als Schriftart (statt computer modern)
  \usepackage{graphicx}                        % um Bilder einzubinden
  \usepackage{floatflt}                        % floating-Umgebung (Textfluss um Bilder oder Tabellen)
  \usepackage{xcolor}                          % fuer farbigen Text etc.
  \usepackage{amsmath}
  \usepackage{amsfonts}
  \usepackage{amssymb}
  \usepackage{bm}
  \usepackage{nicefrac}
  \usepackage{dsfont}
  \usepackage{mathrsfs}
  \usepackage{multirow}                        % um in Tabellen mehrere Felder einer Spalte zusammenzufassen (analog zu multicolumn)
  \usepackage{tabularx}                        % stellt tabularx-Umgebung zur Verfuegung (Tabellen mit fester Breite)
  \usepackage{longtable}                       % stellt longtable-Umgebung, fuer Tabellen, die ueber mehrere Seiten gehen
  \usepackage[breaklinks=true]{hyperref}
  \usepackage{cleveref}
  \usepackage[]{siunitx} % fuer physikalischen Groessen

  \usepackage{lipsum}

\newcommand{\laplacian}{\Delta}

\newcommand{\abs}[1]{\ensuremath{ \left| #1 \right| }}                    % Absolutbetrag einer Zahl |...|
\newcommand{\mean}[1]{\ensuremath{ \left\langle#1\right\rangle }}         % Mittelwert/Erwartungswert eines Ausdrucks <...>

\newcommand{\dd}[0]{\mathrm d}                % d fuer Differential
\newcommand{\del}[0]{\partial}                % partielle Ableitung

\renewcommand{\rho}[0]{\varrho}
\renewcommand{\theta}[0]{\vartheta}
\renewcommand{\phi}[0]{\varphi}
\newcommand{\landauO}[1]{\mathcal{O}\left( #1 \right)}

\renewcommand{\vec}[1]{\bm{#1}}        % Vektoren fett

\newcommand{\ba}[1]{\begin{align} #1 \end{align}}

\newcommand{\muS}{\mu_S}
\newcommand{\kB}{k_\mathrm{B}}
\newcommand{\Neel}{N\'{e}el}

\definecolor{DarkOrange}{RGB}{255,80,0}

\begin{document}

%% ------------------------------------------------------------------- %%
%% --- title, authors, date --- %%
\title{Transport properties of spin superfluids---comparing easy-plane ferro- and antiferromagnets}% Force line breaks with \\

\author{Martin Evers}
\author{Ulrich Nowak}%
\affiliation{Fachbereich Physik, Universit\"at Konstanz, 78457 Konstanz, Germany}%

\date{\today}

%% ------------------------------------------------------------------- %%
%% --- Abstract of paper --- %%
\begin{abstract}
  We present a study on spin-superfluid transport based on an atomistic, classical spin model.
  Easy-plane ferro- as well as antiferromagnets are considered, which allows for a direct comparison of these two material classes based on the same model assumptions. We find a spin-superfluid transport which is robust against variations of the boundary conditions, thermal fluctuations, and dissipation modeled via Gilbert damping.
  Though the spin accumulations is smaller for antiferromagnets the range of the spin-superfluid transport turns out to be identical for ferro- and antiferromagnets.
  Finally, we calculate and explore the role of the driving frequency and especially the critical frequency, where phase slips occur and the spin accumulation breaks down.
\end{abstract}

%% ------------------------------------------------------------------- %%
%% --- create title --- %%
\maketitle

%% %%%%%%%%%%%%%%%%%%%%%%%%%%%%%%%%%%%%%%%%%%%%%%%%%%%%%% %%
%% Introduction
\section{Introduction}

Spin transport in magnetic insulators \cite{Wu13_MagneticInsulators,Nakata17_SpinCurrentsInsulatingMagnets} has been intensively studied beacause of the fundamental interest in the various physical phenomena that occur in these materials and because of their potential for future applications. Magnetic insulators do not exhibit Joule heating \cite{Bauer12_SpinCaloritronics} as no electron transport is involved and many of these are oxides with exceptionally low magnetic damping \cite{Cherepanov93_ReviewYIG}, which hopefully allows for energy efficient transport properties.
It has even been shown that the realization of logic elements is possible \cite{Chumak14_MagnonTransistor}, such that devices are compatible and integratable with CMOS technology \cite{Chumak15_MagnonSpintronics}.
Studies on transport in this material class focuses mostly on transport of magnons \cite{Demokritov13_Magnonics}, i.e.\ quanta of spin waves---the elementary excitations of the magnetic ground state.
As magnons are quasi particles, their number is not conserved and each magnon mode shows an exponential decay upon transport through the system on a length scale $\xi$ called magnon propagation length \cite{Ritzmann14_MagnonPropagation,Zhang12_RelaxationTHzMagnons,Hoffman13_LLTheorySSE,Kehlberger15_LengthScaleSSE,Cornelissen16_MagnonTransportByChemicalPotential,Ritzmann17_SSE_TwoSublMagnets}.
This is even true at zero temperature and in a clean system without any disorder due to the coupling of the magnons to electronic and phononic degrees of freedom, a fact which is described phenomenologically via Gilbert damping in the equation of motion as will be explained below.

In contrast to this damped magnonic transport, a proposal for spin transport was made that carries the name spin superfluidity. The original idea is in fact quite old \cite{Halperin69_HydroTheorySpinWaves,Sonin78_SSF} and rests on a similarity of the magnetic order parameter---either the magnetization of a ferromagnet or the \Neel{} vector of an antiferromagnet---compared to the order parameter of superfluidity---the macroscopic wave function---as it occurs for He-4 below the lambda transition.
For instance, in easy-plane ferromagnets the magnetization features a spontaneously broken rotational symmetry in the easy plane ($SO(2)$ symmetry) that is equivalent to the spontaneously broken gauge invariance of the macroscopic wave function ($U(1)$ symmetry). 
This symmetry leads in both cases to currents that are stable against small deviations---the supercurrents.\ \cite{Sonin17_SSFinYIG}
One striking difference of spin-superfluid transport to spin-wave transport is its distance dependence: for spin superfluidity it is expected to be non-exponential, pushing the limit of the range of magnonic transport.

The first experimental realizations of a spin superfluid was achieved in a system of nuclear spins of He-3 atoms \cite{Bunkov95_SSF_HE3_NMR}---a model system which is not in a solid state. 
Only recently the physics of spin superfluidity has drawn again attention for the case of solid magnets \cite{Koenig01_DissipationlessSpinTransport,Takei14_SSF_EasyPlane,Takei14_SSF_AFM,Flebus16_TwoFluidSSF,Skarsvag15_SSF_Dipole,Yuan18_LongDistTransportCr2O3}, including a proposed dissipationless transport in metallic magnets \cite{Koenig01_DissipationlessSpinTransport}. However, König et al. neglected spin-orbit interaction in their model for the electrons, which is one of the reasons for Gilbert damping in magnets  \cite{Mondal16_RelativisticTheoryElectronRelaxationInLLG}.
But every known material exhibits spin-orbit interaction---since spin and angular momentum of an atom are never exactly zero---and therefore also magnetic damping, even if it is small.
Consequently, spin superfluids do always show dissipation in contrast to their conventional counterparts.

Recent theoretical work has focused on insulators rather than metals, usually based of phenomenological models including the Landau-Lifshitz-Gilbert equation of motion for both ferro- and antiferromagnets. \cite{Takei14_SSF_EasyPlane,Takei14_SSF_AFM,Sonin17_SSFinYIG}
The experimental detection of spin superfluidity in solid-state magnets has been reported for magnon condensates \cite{Bozhko16_MagnonSupercurrent}, where the origin of the spin-superfluid order parameter is different to the cases described above, and also in antiferromagnetic solids \cite{Yuan18_LongDistTransportCr2O3}.
However, the interpretation of the experimental findings is still controversially discussed \cite{Sonin16_CommentsOnMagnonSupercurrents,Bozhko16_CommentsOnMagnonSupercurrents,Sonin17_SSFinYIG,Lebrun18_LongDistTransportHematite}.

\begin{figure}[t!]
  \includegraphics[width=\linewidth]{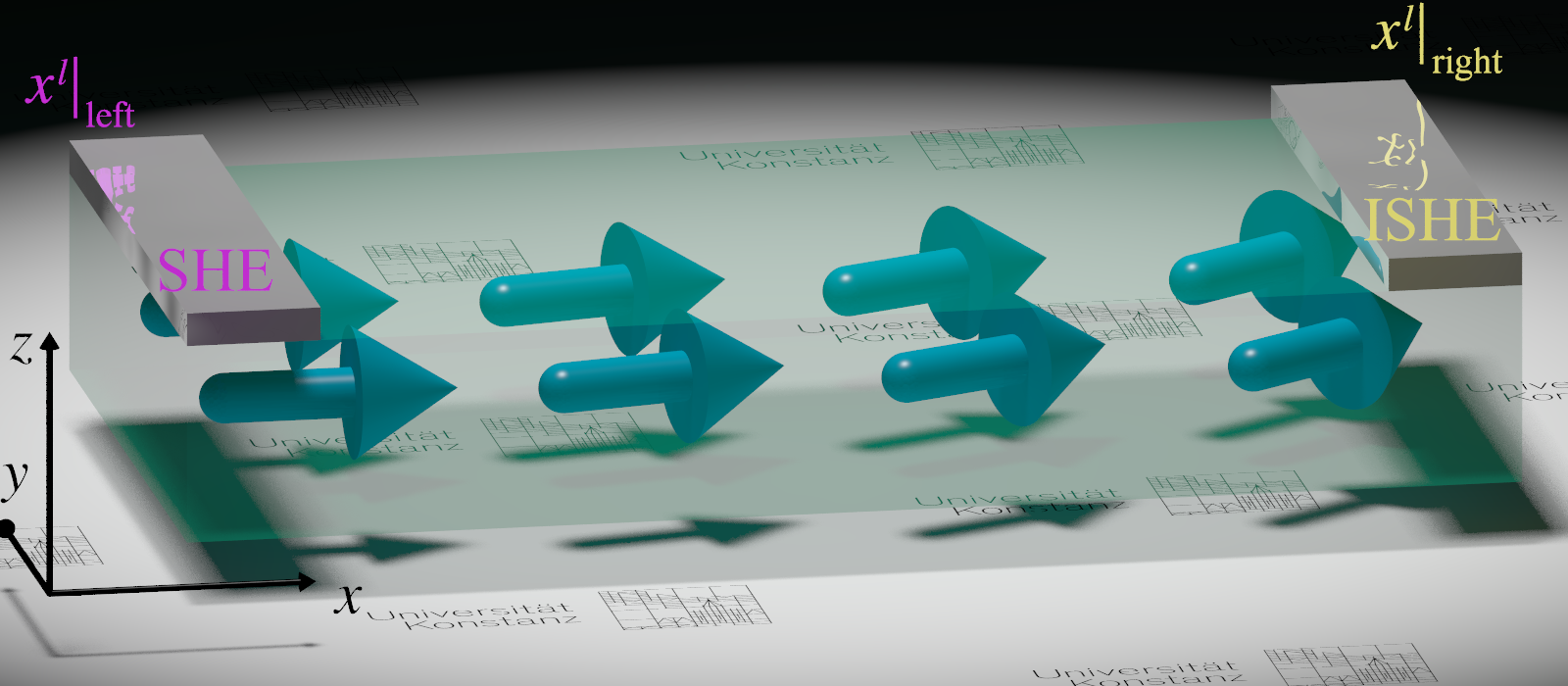}
  \caption{%
    Basic concept of non-local spin transport as in an experimental setup: heavy metal stripes are attached to the magnet to inject a spin current via the spin-Hall effect (here on the left hand side). The spin current in a certain distance (here at the right end) is detected via inverse spin-Hall effect. 
  }
  \label{fig:BasicConcept}
\end{figure}

In the following, we will investigate and compare spin superfluidity in ferro- and antiferromagnetic models. The geometry of our model resembles that of an experimental non-local spin-transport investigation as sketched in \cref{fig:BasicConcept}.
In the corresponding experiments \cite{Cornelissen15_LongDistantTransportYIG_RoomTemp} at one side (here on the left) a spin current is injected into the magnet via the spin-Hall effect caused by an electrical current through an attached heavy-metal stripe.
The resulting spin current is detected using the inverse spin-Hall effect at another position (here the right-hand side).
In our model we avoid the details of the excitation mechanism and model the effect of the injected spin current by an appropriate boundary condition that triggers the dynamics of the spin systems that we investigate.
This is done from the perspective of an atomistic, classical spin model, which has some advantages: the approach is not restricted to small deviations from the ground state, finite temperatures can be investigated and our calculations are not limited to the steady state only.
Furthermore, we are able to compare ferro- and antiferromagnetic systems.
Their behavior turns out to be very similar, except for the resulting spin accumulation that is much lower for the latter.
However, from an experimental point of view antiferromagnets are much more promising, since these are not prone to a breakdown of spin superfluidity as a consequence of dipolar interactions, which is hard to avoid in ferromagnets. \cite{Skarsvag15_SSF_Dipole}

%% %%%%%%%%%%%%%%%%%%%%%%%%%%%%%%%%%%%%%%%%%%%%%%%%%%%%%% %%
%% Model and methods
\section{Atomistic spin model}  \label{sec:ASM}
We consider the following classical, atomistic spin model of Heisenberg type \cite{Nowak07_SpinModels}, comprising $N$ normalized magnetic moments $\vec{S}^l = \boldsymbol{\mu}^l/\muS$ on regular lattice sites $\vec{r}^l$.
We assume a simple cubic lattice with lattice constant $a$.
The Hamiltonian for these moments, in the following called ``spins'', is given by
\ba{
  H = & -\frac{J}{2}\sum_{\langle n,m \rangle}\vec{S}^n{\cdot}\vec{S}^m - d_z\sum_n \left(S_z^n\right)^2,
}
taking into account Heisenberg exchange interaction of nearest neighbors quantified by the exchange constant $J$, where each spin has $N_\mathrm{nb}$ nearest neighbors.
Furthermore, a uni\-axial ani\-sotropy with respect to the $z$ direction with anisotropy constant $d_z$ is included.
In this work we consider the easy-plane case $d_z < 0$, where the magnets ground state reads ${^\mathrm{g}\!}\vec{S}^l = \pm(\cos( {^\mathrm{g}\!\!\:}\phi ), \sin( {^\mathrm{g}\!\!\:}\phi ), 0)$ with some arbitrary, but uniform angle ${^\mathrm{g}\!\!\:}\phi \in [0,2\pi]$ ($SO(2)$ symmetry) and an alternating sign $\pm$ in case of antiferromagnetic order ($J < 0$). 

The time evolution of the spins $\vec{S}^l$ is governed by the stochastic Landau-Lifshitz-Gilbert (LLG) equation of motion \cite{Landau35_LL_equation,Gilbert55_Gilbert_damp,Gilbert04_Gilbert_damp_IEEE}
\ba{
  \frac{\dd \vec{S}^l}{\dd t} & = -\frac{\gamma}{\muS(1+\alpha^2)}\left[ \vec{S}^l \times \left( \vec{H}^l + \alpha\vec{S}^l \times \vec{H}^l \right) \right]  \label{eq:LLG} \\
                    \vec{H}^l & = -\frac{\del H}{\del \vec{S}^l} + \vec{\xi}^l \nonumber \\
        \mean{\xi^l_\beta(t)} & = 0, \; \mean{\xi^l_\beta(t)\xi^{l'}_\eta(t')} = \delta_{ll'}\delta_{\beta\eta}\delta(t-t')\frac{2\muS\alpha\kB T}{\gamma} \nonumber
}
describing the motion of a spin in its effective field $\vec{H}^l$, where $\gamma$ is the gyromagnetic ratio, $\alpha$ the Gilbert damping constant, $\kB$ the Boltzmann constant and $T$ the absolute temperature.
The properties of the thermal noise $\vec{\xi}^l$ are chosen such that the dissipation-fluctuation theorem is satisfied \cite{Brown63_ThermalFluctuactionsMagnParticles}.
The material parameters define our system of units, $|J|$ for the energy, $t_J := \nicefrac{\muS}{\gamma|J|}$ for the time, $a$ for the distance.
Numerically the LLG equation is solved either by the classical Runge-Kutta method in case of zero temperature, or at finite temperature using stochastic Heun's method.
At zero temperature the dissipated power per spin due to Gilbert damping follows directly from the time evolution of the spins $\vec{S}^l(t)$ \cite{Magiera09_SpinRelaxationMagneticTipOnMonolayer}:
\ba{
  P_\mathrm{diss} = \frac{1}{N}\frac{\dd H}{\dd t} = \frac{1}{N}\sum_n \underbrace{\frac{\del H}{\del \vec{S}^n}}_{\mathrm{eff. field}} \cdot \underbrace{\frac{\del \vec{S}^n}{\del t}}_{\mathrm{LLG}}.
  \label{eq:Pdiss}
}
We study a magnetic wire extended along $x$ direction. The system size for our numerical simulations is given by $N = N_x \times N_y \times N_z$ spins along $x$-, $y$- and $z$ direction, where $N_x \gg N_y,N_z$. 
For transverse directions we use periodic boundary conditions if not noted otherwise.
Boundary spins at $x = N_x a$ (the right-hand side) are denoted $\vec{S}^l\big|_\mathrm{right}$ and at this side an open boundary condition is applied, $\vec{S}^l\big|_\mathrm{right} = 0$.
At the opposite side, at $x = 0$, we use a time-dependent boundary condition,
\ba{
  \vec{S}^l\big|_\mathrm{left} = \pm(\cos(\omega_0 t), \sin(\omega_0 t), 0),  \label{eq:TDBC}
}
in form of an in-plane precession with frequency $\omega_0$ that injects a spin current from this side.
The alternating sign ($\pm$) is used only for antiferromagnetic systems, according to the sublattices with antiparallel spin orientation.

The use of this boundary condition creates an excitation with well-defined frequency $\omega_0$.
Alternatively, we also assumed an externally given spin accumulation $\vec{\mu} = \mu\vec{e}_z$ at the left-hand side that causes additional torques on the spins and drives them out of equilibrium, which directly maps an experimental implementation using a spin-Hall-generated spin accumulation to the model utilized here.
This method has been used for instance in \cite{Skarsvag15_SSF_Dipole}.
In \cref{app:ExcitationFrequency} we calculate how this spin accumulation maps to the excitation frequency $\omega_0$ and we furthermore confirmed numerically that both mechanisms lead to the same response for ferro- and antiferromagnets.

Although an atomistic picture---comprising discrete degrees of freedom---is studied numerically, the micromagnetic approximation is of particular value for analytical considerations of ferromagnets.
This approximation assumes that spatial variations of magnetic structures are small compared to the atomic distance $a$.
In this case differences can be approximated as derivatives and the spins form a continuous field $\vec{S}(\vec{r},t)$.
It is handy to use cylindrical coordinates
\[
  \vec{S} = \left( \sqrt{1 - S_z^2}\cos\phi ,\, \sqrt{1 - S_z^2}\sin\phi , \, S_z \right),
\]
where definitions $S_z(\vec{r}^l) := S_z^l$ and $\phi(\vec{r}^l) := \phi^l$ link the atomistic picture to the micromagnetics.
Note that for a spin superfluid $S_z$ is considered as the spin-superfluid density and $\phi$ its phase.
The use of the micromagnetic approximation for ferromagnets allows to reformulate the LLG equation in terms of differential equations for $S_z$ and $\phi$ that read
\begin{widetext}
\ba{
  \frac{\muS}{\gamma} \dot{\phi} & = Ja^2\left[ \frac{1}{1 - S_z^2}\laplacian S_z + \frac{S_z\abs{\nabla S_z}^2}{\left(1 - S_z^2\right)^2} + S_z\abs{\nabla\phi}^2 \right] + 2d_zS_z - \alpha\frac{\muS}{\gamma} \frac{\dot{S_z}}{1 - S_z^2} \label{eq:SSF_FTE_dotphi_2} \\
  \frac{\muS}{\gamma} \dot{S_z}  & = -Ja^2\left[ \left(1 - S_z^2\right)\laplacian\phi -2S_z \nabla S_z \cdot \nabla \phi \right] + \alpha\left(1 - S_z^2\right)\frac{\muS}{\gamma}\dot{\phi}. \label{eq:SSF_FTE_dotSz_2}
}
\end{widetext}
These two equations are strictly equivalent to the LLG equation \cref{eq:LLG} for zero temperature with the only assumption of the micromagnetic approximation.
If one expands these equations in lowest order in $\nabla\phi$, $\laplacian \phi$, $\nabla S_z$, and $\laplacian S_z$ for an easy-plane magnet, which implies especially assuming $|S_z| \ll 1$, but keeping $\abs{\nabla\phi}^2$, one ends up with
\ba{
  \frac{\muS}{\gamma}\dot\phi  & = Ja^2\laplacian S_z + Ja^2S_z\abs{\nabla\phi}^2 + 2d_zS_z -\alpha\frac{\muS}{\gamma}\dot{S_z} \label{eq:HydroEq_dotphi} \\
  \frac{\muS}{\gamma}\dot{S_z} & = -Ja^2\laplacian\phi +\alpha\frac{\muS}{\gamma}\dot\phi \label{eq:HydroEq_dotSz}.
}
Importantly, keeping the $\abs{\nabla\phi}^2$ term is actually required if the damping takes relatively high values, a fact which we checked numerically.
Furthermore, these equations are very similar to others already reported in \cite{Takei14_SSF_EasyPlane,Flebus16_TwoFluidSSF}, but not exactly equivalent.
Ref.~\cite{Takei14_SSF_EasyPlane} uses more approximations, especially neglecting the $|\nabla\phi|^2$-term, and ref.~\cite{Flebus16_TwoFluidSSF} considers a different starting point, namely a quantum theory at low temperatures, where this term has a different $S_z$-dependence.
Because of this difference, the result from \cite{Flebus16_TwoFluidSSF} does not exactly match our numerical results of the atomistic spin model, nor does it match the classical micromagnetic theory.
Hence, we use \cref{eq:HydroEq_dotphi,eq:HydroEq_dotSz} that do describe the atomistic spin simulations well.
However, \cref{eq:HydroEq_dotphi,eq:HydroEq_dotSz} can be solved in steady state for a special case: a ferromagnet that is of length $L$ along $x$ direction and exhibits translational invariance in $y$- and $z$ direction as carried out in \cref{app:FM_analyticalTheory}.
Steady state means a coherent precession of all spins with a frequency $\dot{\phi} = \omega_0$ and a stationary profile $S_z(x)$.
This solution of \cref{eq:HydroEq_dotphi,eq:HydroEq_dotSz} reads:
\ba{
  {^\mathrm{s}\!}\phi(x,t) & = \frac{\alpha}{2}\frac{\muS\omega_0}{\gamma J}\frac{\left(x - L\right)^2}{a^2} + \omega_0 t + \phi_0 \label{eq:SSF_FTE_Sol_sphi} \\
  {^\mathrm{s}\!}S_z(x)    & = \frac{ {^\mathrm{s}\!}S_z(L) }{ 1 + \frac{\muS^2\omega_0^2}{2\gamma^2 J d_z} \alpha^2 \left( \frac{x - L}{a} \right)^2 } , \label{eq:SSF_FTE_Sol_sSz}
}
with a spin accumulation at the right end of the system (at $x = N_xa =: L$) of ${^\mathrm{s}\!}S_z(L) = \nicefrac{\muS\omega_0}{2 \gamma d_z}$, a value which is independent of $L$---one of the striking features of spin superfluidity. Another feature is the monotone increase of $\phi$ which implies the formation of an in-plane spin spiral with winding number $N_\mathrm{w}$, which reads $2\pi N_\mathrm{w} = \int \dd \phi = \phi(L) -\phi(0)$.
Note furthermore, that an open boundary condition at the right end is an assumption that leads to solutions \cref{eq:SSF_FTE_Sol_sphi,eq:SSF_FTE_Sol_sSz}, corresponding to a Neumann condition $\nabla\phi\big|_\mathrm{right} = 0$, which must be justified as a realistic choice.

\begin{figure*}
  \includegraphics[width=0.5\textwidth]{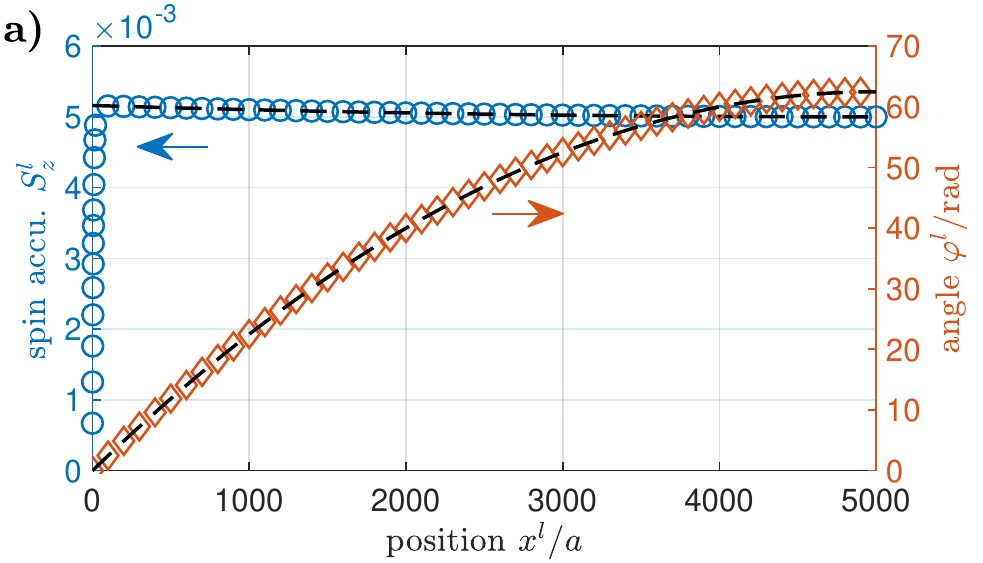}%
  \includegraphics[width=0.5\textwidth]{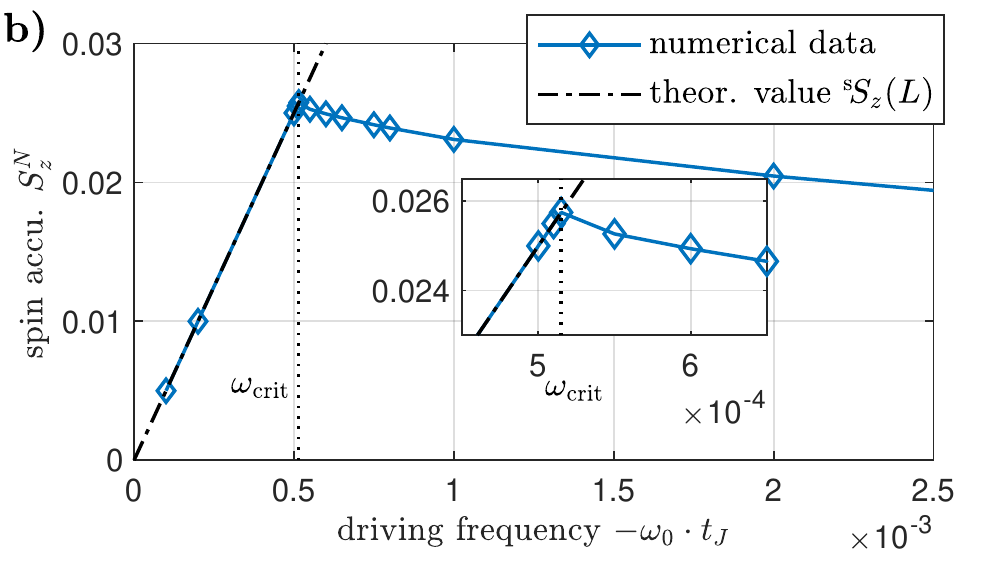}%
  \caption{%
    Spin superfluidity in a 1D ferromagnet at $T = 0$ in the steady state: \textbf{a)} depicts the spin accumulation $S_z$ and the in-plane angle $\phi$ for $\omega_0t_J = \num{-2e-4}$; numerical data (blue and red symbols) follow perfectly the theoretical curve \cref{eq:SSF_FTE_Sol_sphi,eq:SSF_FTE_Sol_sSz} (black, dashed lines), except for the vicinity of the left boundary. This is an artifact of the boundary condition, \cref{eq:TDBC}, used for the numerics. 
    \textbf{b)} shows the spin accumulation at the right end of the system $S_z^N$ versus driving frequency $\omega_0$; for small driving frequencies up to a critical value $\omega_\mathrm{crit}$ the numerical data follow the analytical curve  ${^\mathrm{s}\!}S_z(L)$; for larger frequencies the spin accumulation breaks down, deviating form the theoretical curve, due to phase slips and spin wave excitations.
  }
  \label{fig:FM_T0}
\end{figure*}

For the numerical study of \cref{eq:LLG} we assume an open boundary at the right end.
\Cref{eq:SSF_FTE_Sol_sSz} assumes the same and results in a finite $S_z$ at $x = 0$, which contradicts the numerical driving boundary at this side, \cref{eq:TDBC}, that forces $S_z(x{=}0) = 0$.
Furthermore, in an experiment an open boundary at the right end might not be feasible because of outflowing spin currents, for example into an attached heavy metal. 
Thus, the real behavior at the boundaries for sure deviates from the ideal solution \cref{eq:SSF_FTE_Sol_sSz} and raises the question how strong that deviation is and in how far the boundary conditions influence the overall bulk behavior of the spin transport.
This is examined numerically from the full model \cref{eq:LLG} by varying the boundary conditions on the left and right.
One example of the variations we tested is an absorbing boundary condition on the right, modeling an outflowing spin current by an enhanced damping.
As result we observe the profile $S_z^l$ to show only little change in that case compared to an open boundary and also that in all cases the numerical profiles well follow \cref{eq:SSF_FTE_Sol_sSz} (see in the following \cref{fig:FM_T0} a) as example).
Other variations of the boundary condition which we tested have also hardly any impact on the magnets overall response.

%% %%%%%%%%%%%%%%%%%%%%%%%%%%%%%%%%%%%%%%%%%%%%%%%%%%%%%% %%
%% results on easy-plane ferromagnet

\section{Easy-plane ferromagnet}  \label{sec:FM}
In a first step of the numerical investigation, we consider a collinear ferromagnet as most simple case, with parameters $J > 0$ for the ferromagnetic state and $d_z = -0.01 J$ as in-plane anisotropy.
Let us describe the phenomenology of the spin superfluid in a 1D system of size $N_x \times N_y \times N_z = 5000 \times 1 \times 1$ at temperature $T = 0$. This model is equivalent to a 3D system with translational invariance in $y$- and $z$ direction. Furthermore, we set $\alpha = 0.05$ and $\omega_0 t_J = -\num{2e-4}$.

Starting from a uniform ferromagnet as initial condition, the boundary spin starts to rotate and due to exchange the next spin will follow this rotation and accordingly drive its neighbor and so on.
But since a spin cannot immediately follow the dynamics of its neighbor, there is a certain phase difference $D_\phi$ between the spins, i.e., the neighbor to the right is lagging behind.
In the micromagnetic approximation this effect is described by a phase gradient $\nabla \phi \approx D_\phi / a$.  
The rotation of the spins speeds up, until it reaches the final precession frequency, given by the driving frequency $\omega_0$.
At the same time the out-of-plane component $S_z$---the spin accumulation---increases until it reaches a steady state profile.
The time scale of this transient phase for reaching a steady state can be quantified: $\dot{\phi}(t)$ and $S_z(t)$ follow a limited exponential growth on a characteristic time $\tau_\mathrm{t} \approx \num{5e5}t_J$ for the parameters used here.
$\tau_\mathrm{t}$ scales positively with system size $N_x$ and damping $\alpha$.

Eventually, the numerical time evolution reaches a steady state as shown in \cref{fig:FM_T0} a).
This steady state  verifies the analytical solution \cref{eq:SSF_FTE_Sol_sphi,eq:SSF_FTE_Sol_sSz} in bulk with a deviation only at the left boundary as anticipated and described above.
Note that the finite spin accumulation $S_z$ as a consequence of this type of dynamics has important features: it is a long-range spin transport since it decays non-exponential and it  would allow to measure spin transport by means of the inverse spin-Hall effect.
Furthermore, it could also be addressed, for instance, by magneto-optical measurements---if sensitive to the out-of-plane magnetization for a geometry as studied here.

\begin{figure}
  \includegraphics[width=\linewidth]{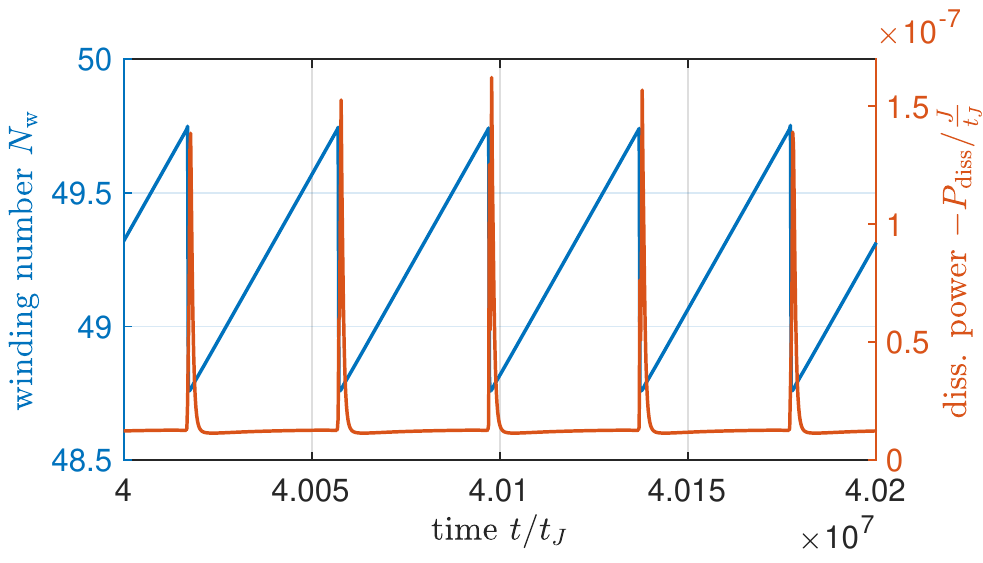}
  \caption{%
    Winding number $N_\mathrm{w}$ and dissipated power $P_\mathrm{diss}$ in the steady state of a driven ferromagnet for $\omega_0t_J = \num{-6.5e-4}$, well in the nonlinear regime. At a rate of $\Gamma_\mathrm{ps}$ phase slips relax the winding number of the in-plane spiral. For each such event the dissipated power spikes.
  }
  \label{fig:FM_Nw_Pdiss_vs_time}
\end{figure}

For a further investigation, we vary the frequency $\omega_0$ and find two different regimes, one for sufficiently small $\omega_0$, where the system is able to follow the excitation without disturbance, and one for large $\omega_0$ where the systems response deviates from the theoretical expectation.
These two regimes, which we will call linear and nonlinear regime in the following, are sharply separated by a critical frequency $\omega_\mathrm{crit}$.
The existence of these two regimes can be seen from the data depicted in \cref{fig:FM_T0} b).
Here, as a measure, we consider the spin accumulation of the last spin $S^N_z$ at the right end of the system.
Below $\omega_\mathrm{crit}$ we find just the analytical value ${^\mathrm{s}\!}S_z(L)$, see \cref{eq:SSF_FTE_Sol_sSz}, which scales linearly with $\omega_0$. At $\omega_\mathrm{crit}$ this behavior breaks down and the spin accumulation $S_z^N$ decreases with increasing pumping frequency.
This breakdown can be understood in terms of the phase gradient $\nabla\phi$ which scales linearly with the driving frequency $\omega_0$, see \cref{eq:SSF_FTE_Sol_sphi}.
However, one can expect a maximum phase gradient $\nabla\phi$ for a spin-superfluid state given by the Landau criterion \cite{Sonin10_SpinCurrentsSSF}: if the phase gradient exceeds locally a critical value, it is energetically favorable for the spins at this position to rotate out of the $x$-$y$ plane and return to the plane by unwinding the spiral.
Hence, the winding number $N_\mathrm{w}$ decreases by one---an effect which is called a phase slip.
The Landau criterion for the stability of a spin superfluid with respect to phase slips reads \cite{Sonin10_SpinCurrentsSSF}
\ba{
  |\nabla\phi| < \sqrt{ -\frac{2d_z}{Ja^2} }. \label{eq:LandauCrit}
}
Note that this relation is not exact as a uniform $S_z$ is assumed for its derivation. Nevertheless do we observe these phase slips numerically.
In the linear regime the winding number is constant in the steady state, whereas in the nonlinear regime it relaxes by one at a regular rate $\Gamma_\mathrm{ps}$ as shown in \cref{fig:FM_Nw_Pdiss_vs_time} at the example of $\omega_0t_J = \num{-6.5e-4}$.

\begin{figure}
  \includegraphics[width=\linewidth]{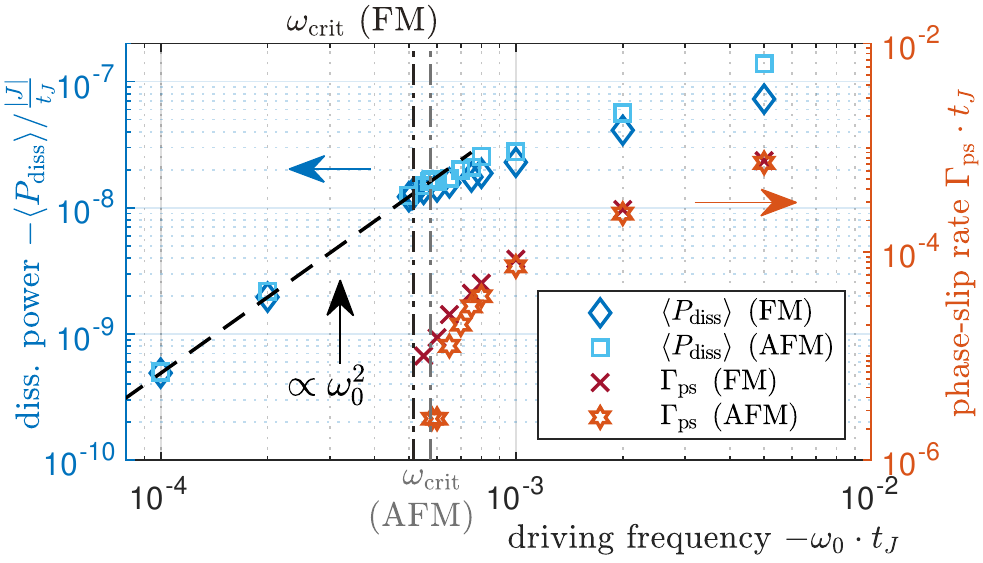}
  \caption{%
    Time-averaged dissipated power $\mean{P_\mathrm{diss}}$ and phase-slip rate $\Gamma_\mathrm{ps}$ in the steady state versus driving frequency $\omega_0$ comparing ferro- (FM) and antiferromagnets (AFM).
    The perpendicular dash-dotted lines mark $\omega_\mathrm{crit}$ for the FM and the AFM, where the latter takes on the higher value.
    For $\omega_0 < \omega_\mathrm{crit}$ the dissipated power scales as $\mean{P_\mathrm{diss}} \propto \omega_0^2$ and is identical for FMs and AFMs.
    Above $\omega_\mathrm{crit}$ the increase slows down and the curve flattens for very high $\omega_0$.
    In this regime, $\mean{P_\mathrm{diss}}$ is higher for AFMs as compared to FMs.
    For the phase-slip rate we find $\Gamma_\mathrm{ps} = 0$ for $|\omega_0| < |\omega_\mathrm{crit}|$ and similar, increasing values for the FM and the AFM above $\omega_\mathrm{crit}$.
    The deviation between FM and AFM near $\omega_\mathrm{crit}$ is due to the different critical frequency, i.e.\ the data almost coincide when plotted versus $\omega_0 - \omega_\mathrm{crit}$.
  }
  \label{fig:FM_Pdiss}
\end{figure}

The $\omega_0$ dependence of the phase-slip rate $\Gamma_\mathrm{ps}$ is depicted in \cref{fig:FM_Pdiss}.
Each phase slip is accompanied by the excitation of a broad spin-wave spectrum on top of the spin superfluid.
These spin waves are visible as oscillations of $S_z$ around the spin-superfluid magnitude and, hence, there is strictly speaking no steady state any more as the phase slips and the spin-wave excitation are definitely time dependent.
In particular, for systems with low Gilbert damping this dynamics leads to deterministic chaos, though the spin-superfluid background remains visible.
These findings have some severe implications as there is a maximum spin accumulation, which is achieved right at the edge between the linear and the nonlinear regime.
Furthermore, driving the system in the nonlinear regime means also to waste energy to the phase slips and the excitation of incoherent spin waves. 

For the parameters here the critical frequency takes the value $\omega_\mathrm{crit}t_J \approx -\num{5.15e-4}$, which is determined from \cref{fig:FM_T0} b).
We also tested different parameters, varying $\alpha$ and $L$ (data not shown), and find that $\omega_\mathrm{crit}$ scales negatively with $\alpha$ and $L$.
Our numerical result can be compared to the analytical prediction above, \cref{eq:LandauCrit}. 
From \cref{eq:SSF_FTE_Sol_sphi} follows that the maximum phase gradient is given by $\nabla{^\mathrm{s}\!}\phi(0) = \nicefrac{\alpha\muS\omega_0 L}{\gamma J a^2}$, which, inserted into \cref{eq:LandauCrit}, implies
\ba{
  |\omega_\mathrm{crit}| = \frac{\gamma J a}{\alpha \muS L}\sqrt{-\frac{2d_z}{J}}.
}
For our parameter set this takes value $\num{4e-4}\nicefrac{1}{t_J}$, which is slightly lower compared to the numerical value above.
This discrepancy is probably due to the fact that \cref{eq:LandauCrit} ignores the spatial dependence of $S_z$.
Furthermore, a test for very low damping $\alpha = \num{e-4}$ showed that \cref{eq:LandauCrit} is even more inaccurate in that case.

Another important quantity is the dissipated power given by \cref{eq:Pdiss}, which takes negative sign as it lowers the total energy.
\Cref{fig:FM_Pdiss} depicts its dependence on $\omega_0$.
Below the critical frequency, in the steady state, it is time independent as the dynamics is completely stationary.
In this regime it scales quadratically with the excitation frequency, $P_\mathrm{diss} \propto \omega_0^2$, a result which has already been reported before \cite{Ochoa18_SSF_AmorphousMagnets}. 
This behavior changes above $\omega_\mathrm{crit}$.
The time evolution of the dissipation in this regime shows that the phase slips notably contribute to the dissipated power, i.e.\ for each phase slip $P_\mathrm{diss}$ spikes as shown in \cref{fig:FM_Nw_Pdiss_vs_time}.
Because of this time-dependence of $P_\mathrm{diss}$, we have to consider an average over time for the evaluation of the dissipated power.
Still, the dissipated power increases further with $\omega_0$ but less than linear and the curve notably flattens. 

%% Finite temperature
So far, our results were obtained from zero-temperature simulations.
In the following we address the robustness of spin-superfluid transport at finite temperature.
For this, we consider a finite cross section $N_y \times N_z > 1$ and $N_x = 2000$ and vary the temperature.
An average over $N_\mathrm{av}$ realizations of thermal noise is carried out and, furthermore, data are averaged over the cross section in order to reduce the noise.
The specific choice of parameters in provided is \cref{tab:FM_FiniteT}. 

\begin{figure}
  \includegraphics[width=0.47\textwidth]{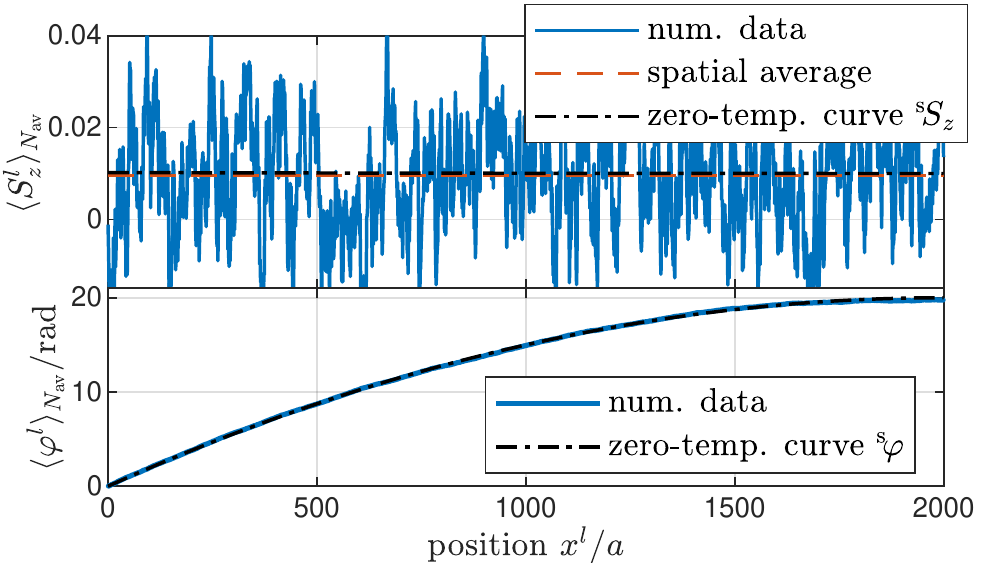}%
  \caption{%
    Spin superfluidity in a ferromagnet at finite temperature $\kB T/J = \num{e-2}$ and for $\omega_0 t_J = -\num{2e-4}$: shown is the spin accumulation $S_z$ and the in-plane angle $\phi$.
    Blue lines represent the numerical data, black dash-dotted lines the analytical solution at zero-temperature.
    The spin accumulation is subjected to strong thermal fluctuations but still has a finite average value $\mean{S_z}_x = \sum_n S_z^n/N_x$, depicted as red dashed line.
    Its value is only slightly lower than the zero-temperature value. Thermal fluctuations are much less pronounced for the in-plane angle.
  }
  \label{fig:FM_finiteT}
\end{figure}

\Cref{fig:FM_finiteT} presents the numerical results for the example of $\kB T/J = \num{e-2}$ for $S_z$ and $\phi$.
The spin-superfluid transport remains in tact but, in particular, the spin accumulation $S_z$ shows strong thermal fluctuations despite the averages taken over the cross section and the $N_\mathrm{av}$ realizations.
However, on average the spin accumulation clearly deviates from its equilibrium value, which is zero.
To quantify the influence of the temperature we calculate the spatial average over the $x$ direction $\mean{S_z}_x$ and compare this to the zero-temperature value, given by \cref{eq:SSF_FTE_Sol_sSz}.
The results are included in  \Cref{tab:FM_FiniteT}.
Furthermore, the in-plane angle $\mean{\phi}_{N_\mathrm{av}}$ shows only little fluctuations and its spatial profile shows hardly any deviation from the zero-temperature behavior, given by \cref{eq:SSF_FTE_Sol_sphi}.
Overall, we find no significant difference to the zero temperature case.
We also checked whether phase slips due to thermal activation can be observed, but from the available data we could not observe a single one with the conclusion that $\Gamma_\mathrm{ps}t_J < \num{4e-5}$. 
Hence, spin superfluidity is very robust against thermal fluctuations, even though these fluctuations are a problem in our simulations in terms of the signal-to-noise ratio.

\begin{table}
  \caption{%
    Averaged spin accumulation of a ferromagnet driven with $\omega_0 t_J = -\num{2e-4}$ for different temperatures.
    The corresponding zero-temperature value is $\mean{S_z} = 0.01$, from which no significant deviation is observed.
  }
  \begin{tabular}{l|l|l|l}
    $\kB T /J$  & $N_x \times N_y$ & $N_\mathrm{av}$ & $\mean{S_z}_x$ \\
    \hline
    \num{e-4}   & $4 \times 4$     & $38$            & \num{0.010} \\
    \num{e-2}   & $4 \times 4$     & $15$            & \num{0.009} \\
    \num{0.05}  & $8 \times 8$     & $5$             & \num{0.010} \\
    \num{0.10}  & $8 \times 8$     & $4$             & \num{0.011} \\
    \num{0.20}  & $14 \times 14$   & $5$             & \num{0.012} \\
  \end{tabular}
  \label{tab:FM_FiniteT}
\end{table}

%% %%%%%%%%%%%%%%%%%%%%%%%%%%%%%%%%%%%%%%%%%%%%%%%%%%%%%% %%
%% results on easy-plane antiferromagnet
\begin{figure*}
  \includegraphics[width=0.50\textwidth]{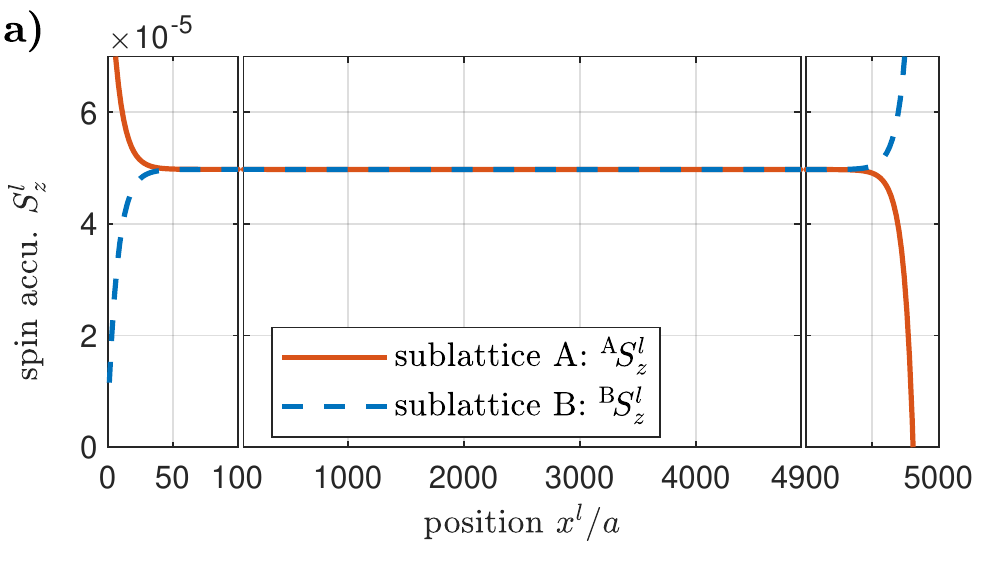}%
  \includegraphics[width=0.50\textwidth]{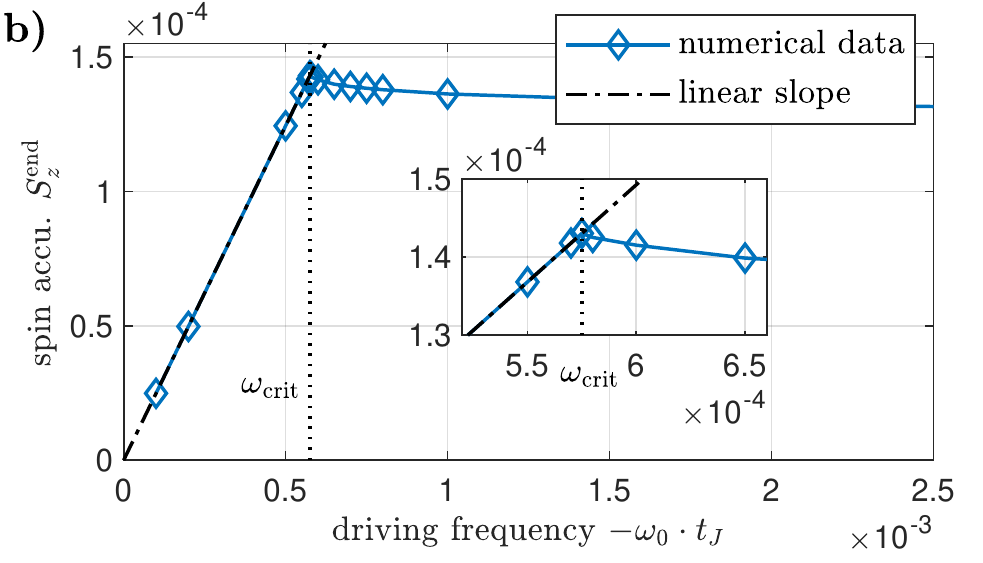}%
  \caption{Spin superfluidity in antiferromagnetic spin chains: \textbf{a)} the spin accumulation in the stead state resolved for the two sublattices A and B.
  In the bulk both take the same value, leading to a finite total spin accumulation, which is two orders of magnitude lower as compared to the ferromagnet.
  At the boundaries the profiles show deviations from bulk behavior because of the broken exchange right at the boundary.
  \textbf{b}) the spin accumulation at the right end of the system as function of driving frequency $\omega_0$; as for the ferromagnet there are two regimes separated by a critical frequency $\omega_\mathrm{crit}$.}
  \label{fig:AFM_Sz}
\end{figure*}

\section{Easy-plane antiferromagnets} \label{sec:AFM}
For antiferromagnets, the magnetic unit cells comprise two atoms---denoted A and B in the following---that form two sublattices. We write all properties using this labeling so that ${^\mathrm{A}\!}\vec{S}^l$ and ${^\mathrm{B}\!}\vec{S}^l$ are spins of the corresponding sublattices.
In the ground state both sublattices have opposite orientation, ${^\mathrm{A}\!}\vec{S}^l = -{^\mathrm{B}\!}\vec{S}^l$.
The field equations, \cref{eq:SSF_FTE_dotphi_2,eq:SSF_FTE_dotSz_2}, do not hold as these require a small in-plane angle difference between two neighboring spins $D_\phi$, which is obviously not true in this case. However, it is reasonable to define phase differences and gradients within each sublattice, i.e.\ ${^\mathrm{A}\!}D_\phi$ as phase difference between a spin of sublattice A and its next-nearest neighbor, which is the nearest neighbor within sublattice A. 
Accordingly, ${^\mathrm{B}\!}D_\phi$ defines the phase difference of sublattice $B$. Assuming sufficiently weak excitation, spatial variations within each sublattice are small such that a micromagnetic approximation inside the sublattices reads $\nabla{^\mathrm{A,B}\!}\phi \approx {^\mathrm{A,B}\!}D_\phi/2a$.
Interestingly, numerical results reveal that the antiferromagnetic system in bulk fulfills field equation \eqref{eq:HydroEq_dotSz}, applied separately to each sublattice.
The other \cref{eq:HydroEq_dotphi} is not valid, as has been reported before \cite{Takei14_SSF_AFM} for a phenomenological model for antiferromagnets.
Consequently, the antiferromagnet is expected to exhibit the same in-plane angle ${^\mathrm{A,B}\!}\phi$ (up to phase difference of $\pi$ between sublattices) as a ferromagnet with corresponding parameters, but not the same spin accumulation ${^\mathrm{A,B}\!}S_z$.

Before we discuss the numerical results in detail, let us first introduce two differences to the ferromagnet that are essential for understanding the following results: the role of exchange and (interlinked with this) the transverse geometry. Just as in a ferromagnet, a spin-superfluid dynamics imposes a finite spin accumulation ${^\mathrm{A,B}\!}S_z$ which, remarkably, carries the same sign for both sublattices leading to a small out-of-plane magnetization.
But this is of course antagonized by the antiferromagnetic exchange that favors antiparallel orientation of all components between sublattices.
Consequently, the exchange interactions must lower the spin accumulation $S_z$ tremendously as compared to the ferromagnet (compare \cref{fig:AFM_Sz} a) and \cref{fig:FM_T0} a)).
This also implies that the behavior of $S_z$ is determined by the number of nearest neighbors $N_\mathrm{nb}$ of a spin as more neighbors imply stronger exchange coupling. Consequently, a 1D spin chain is less prone to this exchange reduction than a 3D system. We checked this numerically by comparing 1D, 2D and 3D models and, indeed, the spin accumulation of the spin superfluid $S_z$ scales linearly with $N_\mathrm{nb}$. 

There is another implication: at a boundary the number of neighbors is locally reduced---and therefore the importance of the exchange---, resulting in deviations of the sublattice components ${^\mathrm{A,B}\!}S_z$, see \cref{fig:AFM_Sz} a) for a 1D setup (the effect is less pronounced in 3D).
This 1D setup owns only boundaries along the $x$ direction and the question whether for finite cross section $N_y \times N_z > 1$ these deviations at $y$- and $z$ boundaries significantly influence the bulk behavior has also been tested numerically.
Fortunately, deviations at transversal boundaries quickly fall off with distance to the boundary over a few lattice constants.
The bulk then behaves qualitatively and quantitatively just as a 1D system, except for the reduced spin accumulation due to the number of neighbors as discussed above.
The study of 1D systems is preferable to keep computational costs feasible.

We turn now to the presentation of the numerical data for a 1D system.
The model parameters are the same as given above for the ferromagnet, except for the exchange constant which is now negative.
Similarly to the ferromagnet, the system reaches a steady state after a transient phase characterized by a limited exponential growth on a time scale $\tau_\mathrm{t}$, which is roughly the same as for the ferromagnet.
In the steady state the sublattice-resolved in-plane angles ${^\mathrm{A,B}\!}\phi$ both follow exactly the same profile as the ferromagnet, i.e.\ \cref{eq:SSF_FTE_Sol_sphi}, but with a phase difference of $\pi$ between the two sublattices because of the antiferromagnetic order (data for the antiferromagnet not shown).

The spin accumulation deviates from the behavior of a ferromagnet as depicted in \cref{fig:AFM_Sz} a).
The bulk profiles (away from boundaries at $x = 0$ and $x = N_xa$) are identical in the two sublattices, ${^\mathrm{A}\!}S_z = {^\mathrm{B}\!}S_z$.
Hence, a measurable spin accumulation is present, but it is two orders of magnitude lower than in comparable ferromagnetic cases.
This is the aforementioned exchange reduction. 
If we consider the spin accumulation $S_z$ in bulk, in the data in \cref{fig:AFM_Sz} a) hardly a space dependence is observed in contrast to the ferromagnet, where $S_z^l$ has a finite slope.
The antiferromagnet exhibits this in the same way, but it is also much smaller and the profile becomes roughly constant.
Contrary to the ferromagnet, there are disturbances at the boundaries in the profile of $S_z$ which we already discussed before.

Driving the antiferromagnet with the time-dependent boundary condition \cref{eq:TDBC} at frequency $\omega_0$ leads to the very same two different regimes as for ferromagnets, a linear regime up to a critical frequency $\omega_\mathrm{crit}$ and above---in the nonlinear regime---phase slips occur.
These phase slips reduce the winding number, lead to the excitation of spin waves, and a further increase of the spin accumulation is not possible.
We quantify this behavior in a similar way as for the ferromagnet.
It is, however, not possible to use the spin accumulation of the last spin $S^N_z$ as a measure because of the deviating profile at the boundary.
Instead, we take the spin accumulation at the end of the bulk in form of a spatial average over the spins in the range $x^l/a \in [4900, 4920]$, $S_z^\mathrm{end} := \mean{ S_z^l }_{[4900, 4920]}$.
This range is chosen such that it is sufficiently separated from the boundary.
The data for the $\omega_0$ dependence of the spin accumulation are shown in \cref{fig:AFM_Sz}, panel b): These show that critical frequencies takes roughly same values for ferro- and antiferromagnets, a result which has been tested and confirmed for another parameter set with different $N_x$, $\alpha$, and $d_z$.
For the data set shown here the value is $\omega_\mathrm{crit} t_J \approx \num{-5.75e-4}$.
However, the decrease of the spin accumulation $S_z^\mathrm{end}$ with increasing driving frequency $\omega_0$ in the nonlinear regime is less pronounced for antiferromagnets.
We also calculated the $\omega_0$ dependence of the time-averaged dissipated power $\mean{P_\mathrm{diss}}$ and of the phase-slip rate $\Gamma_\mathrm{ps}$, both shown in \cref{fig:FM_Pdiss}.
Similar to other features these properties behave for the antiferromagnets very much as for ferromagnets: below $\omega_\mathrm{crit}$ the dissipated power shows exactly the same dependence and above it is dominated by phase slips.
However, a difference is that above $\omega_\mathrm{crit}$ the dissipated power increases faster with $\omega_0$.
One reason for this might be the dynamics of spin waves that very much differ between ferro- and antiferromagnets.
The phase-slip rate differs slightly, however, this seems to be solely due to the fact that $\omega_\mathrm{crit}$ differs for ferro- and antiferromagnets.
When $\Gamma_\mathrm{ps}$ is plotted versus $\omega_0 - \omega_\mathrm{crit}$, both curves match almost.

The next step is to consider finite temperature.
Again this requires a finite cross section for which we use $N_x \times N_y \times N_z = 2000 \times 4 \times 4$ and we test two temperatures, $\kB T/J = \num{e-2}$ and $\kB T/J = \num{e-4}$.
As before, the magnetic response is very similar to that of a ferromagnet: the in-plane angles follow the zero-temperature profiles, as well as does the average spin accumulation for the lower of the two temperatures.
The only major difference is the ratio of the spin-superfluid spin accumulation to the thermal fluctuations, which is two orders of magnitude smaller as a result of the lower spin-superfluid signal and an equal strength of the fluctuations.
For the higher temperature, this even leads to an average $S_z$ that is essentially zero.
This does not mean that there is no spin-superfluid spin accumulation, but rather that the available numerical data are not sufficient to resolve it and more averaging is needed.
Note that the in-plane angle is not affected by this---it is as robust against the fluctuations just as for the ferromagnet.

%% %%%%%%%%%%%%%%%%%%%%%%%%%%%%%%%%%%%%%%%%%%%%%%%%%%%%%% %%
%% conclusion
\section{Discussion and Conclusion}
Our comparative study addresses spin superfluidity in ferro- and antiferromagnets, where one should bear in mind that the former are less promising for spin superfluidity as the latter because of the negative influence of the stray field \cite{Skarsvag15_SSF_Dipole}.
Nevertheless, the former can help to understand the behavior of the latter, which we utilize in this work.
One of the striking features of spin superfluidity is the transport range that leads to a spin accumulation at the end of the system $S_z(L)$ (see \cref{eq:SSF_FTE_Sol_sphi,eq:SSF_FTE_Sol_sSz}) that does depend on the system length $L$---a completely different situation compared to spin-wave transport where the intensity decays exponentially with the distance.
However, this non-exponential decay does not imply the possibility of an infinite transport range since with increasing system size the critical frequency lowers until no undisturbed spin superfluid is possible anymore.

We present a full analytical solution for the steady state of the ferromagnet, which slightly deviates from the analytical theory reported before \cite{Takei14_SSF_EasyPlane,Flebus16_TwoFluidSSF}.
This theory is tested numerically by the full atomistic model, which allows to test the robustness of the spin-superfluid transport against varying boundary conditions, against high excitation frequencies and finite temperature.
We show that this kind of transport is remarkably robust: boundary conditions and also elevated temperature hardly hamper the magnets spin-superfluid response.

Furthermore, we identify the critical frequency $\omega_\mathrm{crit}$---a manifestation of the Landau criterion---as the limiting factor for the range of this transport.
Above this critical frequency phase slips occur, which also sets a limit to the spin accumulation that can be achieved.
In ref.~\cite{Tserkovnyak17_CoherenceSSF} another limitation on the spin current of such a spin superfluid is discussed, which rests on the fact that $|S_z|$ is bounded above.
But the estimated values would require an out-of-plane component that takes quite large values $|S_z| > 0.1$, which our simulations reveal to be hardly possible even for low damping.
This is in particular true for the case for antiferromagnets and, therefore, we conclude that the critical frequency---and therefore the phase slips---is a more relevant limitation on spin superfluid transport.

The direct comparison of antiferromagnets to ferromagnets shows that both exhibit the very same behavior:
Driven by an in-plane rotation, both form an in-plane spin spiral that exhibits exactly the same behavior, including a spin accumulation in form of an out-of-plane magnetization.
Antiferromagnets show in principle the same transport range as ferromagnets with a spin accumulation at the end of the system independent of the system length, provided the excitation frequency $\omega_0$ is kept constant ($\omega_0$ itself depends on the magnets geometry in experimental setups, see \cref{eq:app_omega0_spec3D}).
Furthermore, the critical frequency takes very similar value for the two types of magnets.
This general accordance of spin superfluidity for both types of magnets is in contrast to spin-wave transport that is known to be different for ferro- and antiferromagnets \cite{Keffer53_CompareSW_FMs_AFMs}.
Yet there is a major deviation: the antiferromagnetic exchange lowers tremendously the spin accumulation.

Our study also covers an examination of the dissipation of a spin superfluid and of the effect of finite temperature.
We proof the principle robustness of spin superfluidity against thermal fluctuations, i.e.\ that quite high temperatures are required before thermal phase slips start to hamper the transport.
But the fluctuations are a problem from the numerical side as these require integration over a large amount of data in order to identify a non-zero mean spin accumulation.
The signal-to-noise ratio might be a problem in experimental setups as well and it could be more promising to measure rather the in-plane angle $\phi$, which is more robust against thermal fluctuations and which always delivers a clear signal in the cases we investigated here.
A measurement of $\phi$ can be done in two ways: either by its time evolution, i.e.\ the precession frequency $\omega_0$, or spatially resolved by measuring the formation of the in-plane spin spiral.

\acknowledgments{
  Financial support by the Deutsche Forschungsgemeinschaft (DFG) via the SFB 767 ``Controlled Nanosystems: Interaction and Interfacing to the Macroscale'' and the program ``Hematite: A new paradigm for antiferromagnetic spin transport'' is gratefully acknowledged.
}

\bibliography{Bibliography}

\appendix

\section{Analytical theory for a 1D ferromagnet}  \label{app:FM_analyticalTheory}
The ferromagnet in the micromagnetic approximation under the assumption of small out-of-plane component, $|S_z| \ll 1$, is described by the LLG equation in cylindrical coordinates, \cref{eq:HydroEq_dotphi,eq:HydroEq_dotSz}.
Assuming translational invariance along $y$- and $z$ direction leads to a 1D problem:
\ba{
  \frac{\muS}{\gamma} \dot{\phi} & = Ja^2\del_x^2 S_z + Ja^2 S_z (\del_x\phi)^2 + 2d_zS_z - \frac{\alpha \muS}{\gamma}\dot{S_z} \\
  \frac{\muS}{\gamma} \dot{S_z}  & = -Ja^2\del_x^2 \phi + \frac{\alpha \muS}{\gamma}\dot{\phi} .
}
Steady state means $\dot{\phi} = \omega_0$ and $\dot{S_z} = 0$.
This allows to integrate the latter equation,
\ba{
  {^\mathrm{s}\!}\phi(x,t) = \frac{\alpha}{2}\frac{\muS \omega_0}{\gamma J} \left( \frac{x - L}{a} \right)^2 + \omega_0t + \phi_0,
  \label{eq:AppA_Solphi}
}
where the first integration constant follows from the Neumann boundary condition at the right end, $\del_x\phi(L) = 0$ (no outflow of spin current), and the second one satisfies the condition $\dot{\phi} = \omega_0$ and allows for an arbitrary phase $\phi_0$.
This is inserted in the first equation, which then reads
\ba{
  -Ja^2 \del_x^2 S_z = - \frac{\muS \omega_0}{\gamma} + \frac{\muS^2 \omega_0^2}{\gamma^2 J} \left( \alpha \frac{x - L}{a} \right)^2S_z + 2d_z S_z.  \label{eq:AppA_HydroEq_dotphi_test_LaplacianSz}
}
We argue that the second-derivative term can be neglected $-Ja^2 \del_x^2 S_z \approx 0$.
This is justified in a twofold manner: first we compared the relevance of all terms in that equation numerically by calculating those three terms from simulations of the full atomistic spin model, \cref{eq:LLG}.
Indeed the result is that in steady state the second-derivative term is several orders of magnitude smaller compared to the other two. 
The second reason follows a-posteriori from the calculated solution and is explained below.
From $-Ja^2 \del_x^2 S_z \approx 0$ follows the steady-state solution for $S_z$:
\ba{
  {^\mathrm{s}\!}S_z = \frac{\frac{\muS\omega_0}{2\gamma d_z}}{1 + \frac{\muS^2\omega_0^2}{2\gamma^2 J d_z} \left(\alpha\frac{x-L}{a}\right)^2}.
  \label{eq:AppA_SolSz}
}

This solution does not fulfill \cref{eq:AppA_HydroEq_dotphi_test_LaplacianSz}, however, we can insert it and calculate the deviation by calculating
\ba{
  \del_x^2 {^\mathrm{s}\!}S_z & = - 2\frac{\muS\omega_0}{\gamma J} \frac{\alpha^2}{a^2} {^\mathrm{s}\!}S_z^2
                                  + 4\left(\frac{\muS\omega_0}{\gamma J}\right)^2 \frac{\alpha^4(x-L)^2}{a^4} {^\mathrm{s}\!}S_z^3  \nonumber \\
                              & = \landauO{S_z^2}.  \nonumber
}
This allows the conclusion that the correction by taking the second derivative into account is of higher order in $S_z$ and neglecting this is consistent with the original assumption $|S_z| \ll 1$.
Hence, \cref{eq:AppA_Solphi,eq:AppA_SolSz} form the analytical solution for a 1D setup.

\begin{figure}
  \includegraphics[width=\linewidth]{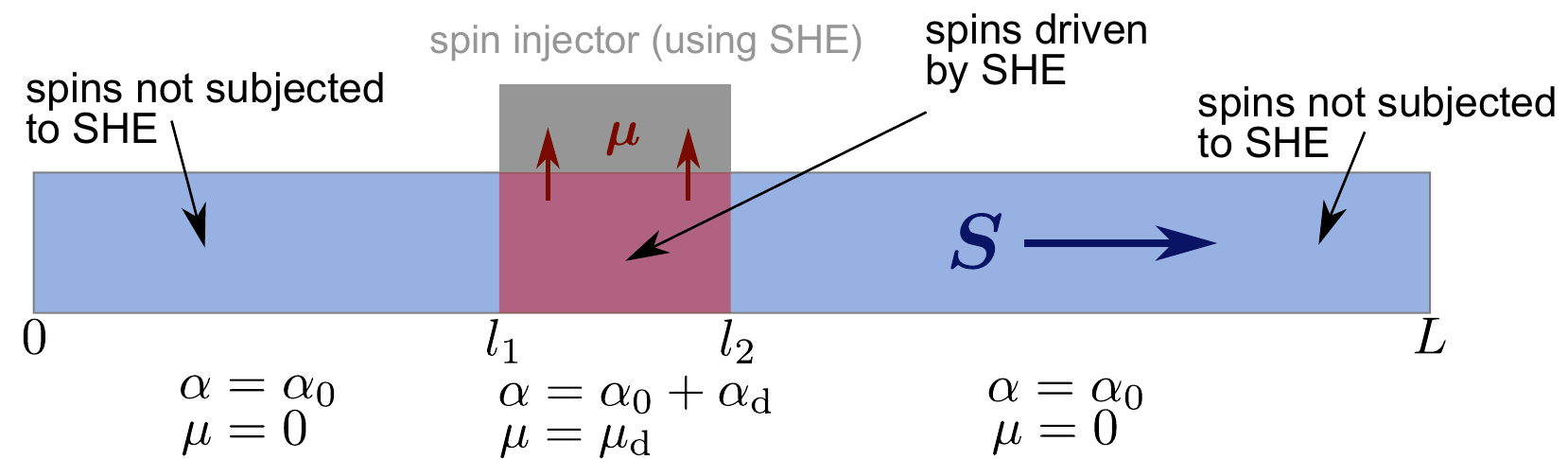}
  \caption{%
    1D setup for calculation of the excitation frequency $\omega_0$ of a magnet driven by a spin injector utilizing the spin-Hall effect to exert external torques on the spins.
    These torques are applied in the region $[l_1, l_2]$ and vanish outside.
    Furthermore, the Gilbert damping in $[l_1, l_2]$ is enhanced by $\alpha_\mathrm{d}$. 
    The ground state $\vec{S}$ is in-plane, the spin accumulation $\vec{\mu}$ perpendicular.
  }
  \label{fig:app_Sketch1DGeometry}
\end{figure}

\section{Frequency of a spin superfluid}  \label{app:ExcitationFrequency}
The usual excitation of a spin current in a magnet rests on a spin accumulation $\vec{\mu}$ at an interface between the magnet and a heavy metal, which is created by an electrical current.
Normally for that the spin-Hall effect is utilized.
The aim of this appendix is to calculate the resulting excitation frequency $\omega_0$ of a spin superfluid.

We assume here that the spin accumulation is perpendicular to the magnets ground state, i.e.\ $\vec{\mu} \propto \vec{e}_z$.
Consequently, there is an additional damping-like torque \cite{Skarsvag15_SSF_Dipole,Manchon19_CurrentInducedSOT_FMs_AFMs} in the LLG equation (here written as viscous damping):
\ba{
  \dot{\vec{S}}^l = -\frac{\gamma}{\muS} \vec{S}^l \times \vec{H}^l + \alpha_l \vec{S}^l \times \dot{\vec{S} }^l + \alpha'_l \vec{S}^l \times \left( \vec{S}^l \times \frac{\vec{\mu}^l}{\hbar} \right) .  \label{eq:app_SHE_LLG}
}
A subset $V_\mathrm{d}$ of the total volume of the magnet is driven, i.e.\ subjected to the additional torques and the driving also creates an enhanced damping $\alpha'_l$ within $V_\mathrm{d}$:
\ba{
  \vec{\mu}^l & = \left\{ \begin{array}{ll}
                            \mu_\mathrm{d}\vec{e}_z & \text{for } \vec{r}^l \in V_\mathrm{d} \\  
                            0 & \text{else}
                          \end{array} \right. \\
  \alpha_l    & = \alpha_0 + \alpha'_l \quad\text{with}\quad  \alpha'_l = \left\{ \begin{array}{ll}
                                                                                     \alpha_\mathrm{d} & \text{for } \vec{r}^l \in V_\mathrm{d} \\
                                                                                     0 & \text{else}
                                                                                  \end{array} \right. .
}
$\alpha_0$ is the intrinsic Gilbert damping of the magnet.

To proceed we consider the LLG equation in the following form, resolved for the time derivative:
\ba{
  \dot{\vec{S}^l} = & -\frac{\gamma}{\muS(1+\alpha_l^2)} \vec{S}^l \times \left( \vec{H}^l + \alpha_l \vec{S}^l \times \vec{H}^l \right)  \nonumber \\
                    & + T_1^l \vec{S}^l \times \vec{A}^l + T_2^l \vec{S}^l \times \left( \vec{S}^l \times \vec{A}^l \right). \label{eq:app_AnsatzLLG_arbTorque} 
}
$T_1^l$ and $T_2^l$ parameterize arbitrary additional torques with respect to an axis $\vec{A}^l$ and for the specific choice $\vec{A}^l = \nicefrac{\vec{\mu}^l}{\hbar}$, $T_1^l = \nicefrac{\alpha_l\alpha'_l}{(1+\alpha_l^2)}$ and $T_2^l = \nicefrac{-\alpha'_l}{(1 + \alpha_l^2)}$ \cref{eq:app_AnsatzLLG_arbTorque} is equivalent to \cref{eq:app_SHE_LLG}.
However, for the sake of generality we consider for the calculation \cref{eq:app_AnsatzLLG_arbTorque}.
Assuming $\vec{A}^l \propto \vec{e}_z$ and using cylindrical coordinates and again the micromagnetic approximation, this form of the LLG reads
\ba{
  \frac{\muS}{\gamma}\dot{\phi} = & \phantom{+} Ja^2S_z|\nabla\phi|^2 + 2d_zS_z - \alpha\frac{\muS}{\gamma}\dot{S_z}  \nonumber \\
                                  & - \frac{\muS}{\gamma}A_z(T_1 + \alpha T_2) \label{eq:appB_eq_phi_withExtTorque} \\
  \frac{\muS}{\gamma}\dot{S_z}  = & -Ja^2 \laplacian\phi + \alpha\frac{\muS}{\gamma} \dot{\phi} +\frac{\muS}{\gamma}A_z(\alpha T_1 - T_2), \label{eq:appB_eq_Sz_withExtTorque}
}
an extension of \cref{eq:HydroEq_dotphi,eq:HydroEq_dotSz}.
In the same spirit as in \cref{app:FM_analyticalTheory} we can solve these equations in one dimension in steady-state (assuming $\dot{S_z} = 0$ and $\dot{\phi} = \omega_0$), where the geometry depicted in \cref{fig:app_Sketch1DGeometry} is assumed.
We apply the external spin accumulation in the interval $[l_1,l_2]$, whereas the total magnet expands over $[0, L]$.
Therefore,
\ba{
  T_{1,2}(x) & = \left\{ \begin{array}{ll}
                          T^\mathrm{d}_{1,2} & \text{ for } x \in [l_1,l_2] \\
                          0 & \text{ else}
                        \end{array} \right. \nonumber \\
  \vec{A}(x) & = \left\{ \begin{array}{ll}
                          A^\mathrm{d}_z\vec{e}_z & \text{ for } x \in [l_1,l_2] \\
                          0 & \text{ else}
                        \end{array} \right. \nonumber .
}

\begin{widetext}
In the 1D setup \cref{eq:appB_eq_Sz_withExtTorque} reads
\ba{
  \del_x^2 \phi & = \alpha(x)\frac{\muS}{\gamma Ja^2} \omega_0 + \frac{\muS}{\gamma Ja^2}A_z(x)\left[ \alpha(x) T_1(x) - T_2(x) \right] \nonumber \\
                & = \left\{ \begin{array}{ll}
                              \overbrace{ \alpha_0\frac{\muS}{\gamma Ja^2} \omega_0 }^{ =:\bar{\omega}_0 } & \text{ for } x \in [0,l_1] \\
                              \underbrace{ (\alpha_0 + \alpha_\mathrm{d}) \frac{\muS}{\gamma Ja^2} \omega_0 }_{ =: \bar{\omega}_0'} 
                                           + \underbrace{ \frac{\muS}{\gamma Ja^2}A_z^\mathrm{d}\left[ (\alpha_0 + \alpha_\mathrm{d}) T_1^\mathrm{d} - T_2^\mathrm{d} \right] }_{ =:t }  &  \text{ for } x \in [l_1,l_2] \\
                              \alpha_0\frac{\muS}{\gamma Ja^2} \omega_0  &  \text{ for } x \in [l_2,L]
                            \end{array} \right. ,
}
which can be integrated.
There are six boundary conditions to consider, each one at the left and right end of the magnet, where we assume a Neumann condition $\del_x \phi(0) = \del_x\phi(L) = 0$, i.e.\ no outflow of spin currents.
Furthermore, $\phi$ and $\del_x\phi$ must be continuous at $l_1$ and $l_2$, delivering four internal boundary conditions.
But there is another condition, a gauge condition for $\phi$, which allows to add an arbitrary constant phase to $\phi(x)$ without altering the physics.
(In practice this gauge phase depends on the prehistory of the magnet, i.e.\ on how it had reached its steady state, and also which exact instant in time is considered.) 
As gauge we use $\phi(0) = 0$.
Altogether there are 6 integration constants and the unknown frequency $\omega_0$ in combination with 6 boundary conditions and a gauge, such that the problem has a unique solution.

As result we obtain
\renewcommand{\arraystretch}{1.25}
\ba{
  \phi & = \left\{ \begin{array}{ll}
                     \frac{1}{2}\bar{\omega}_0x^2 & \text{ for } x\in[0,l_1] \\
                     \frac{1}{2}(\bar{\omega}_0' + t)x^2 + (\bar{\omega}_0 - \bar{\omega}_0' -t)l_1x + \frac{1}{2}(\bar{\omega}_0'-\bar{\omega}_0+t)l_1^2 & \text{ for } x\in[l_1,l_2] \\
                     \frac{1}{2}\bar{\omega}_0x^2 + (\bar{\omega}_0' - \bar{\omega}_0 + t)\left[ (l_2 - l_1)x + \frac{1}{2}(l_1^2 - l_2^2) \right] & \text{ for } x\in[l_2,L]
                   \end{array} \right. \\
  S_z  & = \frac{\muS}{\gamma} \frac{\omega_0 + A_z(x) \left[T_1(x) + \alpha(x)T_2(x) \right]}{Ja^2 (\del_x \phi)^2 + 2d_z}
}
\renewcommand{\arraystretch}{1.0}
and, importantly, we also gain
\ba{
  \omega_0 = -A^\mathrm{d}_z \frac{\left[ (\alpha_0 + \alpha_\mathrm{d})T^\mathrm{d}_1 - T^\mathrm{d}_2 \right](l_2 - l_1)}{\alpha_0 L + \alpha_\mathrm{d}(l_2 - l_1)}. \label{eq:app_omega0_gen}
}
\end{widetext}
This holds true for arbitrary torques taking form \cref{eq:app_AnsatzLLG_arbTorque}.
If the specific case of the spin injector utilizing the spin-Hall effect is considered, then inserting the parameters $T_1$, $T_2$ and $\vec{A}$ reads
\ba{
  \omega_0 = -\underbrace{ \frac{\mu_\mathrm{d}}{\hbar}\alpha_\mathrm{d} }_{ =: \tau } \cdot \frac{l_2 - l_1}{\alpha_0L + \alpha_\mathrm{d}(l_2 - l_1)}. \label{eq:app_omega0_spec1D}
}
The former factor $\tau$ is the strength of the spin-Hall effect on the magnet \cite{Manchon19_CurrentInducedSOT_FMs_AFMs}:
\[
  \tau = \frac{\gamma}{M_\mathrm{s}}\frac{\hbar}{2e}\eta \theta_\mathrm{SH} j_\mathrm{el}\frac{1}{d},
\]
with spin transparency of the interface $\eta$, spin-Hall angle $\theta_\mathrm{SH}$, saturation magnetization $M_\mathrm{s}$ and thickness $d$ of the magnet.
$j_\mathrm{el}$ is the electric current density.
The latter factor in \cref{eq:app_omega0_spec1D} is a geometric factor that is basically the ratio between the driven volume $l_2 - l_1$ and the total volume $L$, weighted with the total damping of the magnet, where the Gilbert-damping enhancement can be expressed as \cite{Skarsvag15_SSF_Dipole}
\[
  \alpha_\mathrm{d} = g^\perp\frac{\hbar^2}{2e^2}\frac{\gamma}{M_\mathrm{s}d},
\]
with transverse spin mixing conductance $g^\perp$ of the interface.
This rigorous derivation holds only true for 1D ferromagnets, however, the natural extension to 2D and 3D is given by
\ba{
  \omega_0 = -\tau \cdot \frac{V_\mathrm{d}}{\alpha_0V + \alpha_\mathrm{d}V_\mathrm{d}}, \label{eq:app_omega0_spec3D}
}
where $V$ is the magnets total and $V_d$ the driven volume.
The validity of this expression has been checked numerically for 1D and 2D systems using various geometries by investigating the full atomistic LLG \cref{eq:app_AnsatzLLG_arbTorque}.
As a result we obtain very good agreement with the analytical calculation except for two cases.
First, when the assumption $|S_z| \ll 1$ is violated and second if the setup is not effectively one dimensional, i.e.\ if the system is not driven over the entire transverse width.
However, such a mismatch in usually small for realistic experimental setups.
We furthermore did not only simulate ferromagnets, but also antiferromagnets with same parameters except for the sign of $J$. 
These simulations result in exactly the same frequencies $\omega_0$ as the corresponding ferromagnets and thus \cref{eq:app_omega0_gen,eq:app_omega0_spec1D,eq:app_omega0_spec3D} are also valid for antiferromagnets, even though note that the resulting spin accumulation deviates.

\end{document}